\documentclass[12pt]{article}
\usepackage{amsmath,amssymb,amsfonts}
\usepackage{color,graphicx,cite,soul}
\usepackage{array,booktabs,longtable}
\usepackage{caption,microtype}

\graphicspath{{figs/}}

\evensidemargin \oddsidemargin
\renewcommand{\arraystretch}{1.2}

\allowdisplaybreaks
\newcommand{\url}[1]{{\tt #1}}

\newcommand{\gmt}{\ensuremath{(g-2)_\mu}}
\newcommand{\br}{{\rm BR}}
\newcommand{\bsg}{BR($b \to s \gamma$)}

\newcommand{\btn}{BR($B_u \to \tau \nu_\tau$)}

\newcommand{\bmm}{\ensuremath{\br(B_s \to \mu^+\mu^-)}}

\newcommand{\ssi}{\ensuremath{\sigma^{\rm SI}_p}}

\newcommand{\Mh}{\ensuremath{M_h}}

\newcommand{\MA}{\ensuremath{M_A}}

\newcommand{\msusy}{M_{\rm SUSY}}
\newcommand{\mgl}{\ensuremath{m_{\tilde g}}}
\newcommand{\msq}{\ensuremath{m_{\tilde q}}}
\newcommand{\msqone}{\ensuremath{m_{\tilde q_1}}}
\newcommand{\msqtwo}{\ensuremath{m_{\tilde q_2}}}

\newcommand{\mstop}[1]{\ensuremath{m_{\tilde t_{#1}}}}
\newcommand{\msbot}[1]{\ensuremath{m_{\tilde b_{#1}}}}
\newcommand{\msqt}{\ensuremath{m_{\tilde q_3}}}
\newcommand{\msl}{\ensuremath{m_{\tilde l}}}

\newcommand{\cha}[1]{\tilde \chi^\pm_{#1}}
\newcommand{\champ}[1]{\tilde \chi^\mp_{#1}}
\newcommand{\mcha}[1]{\ensuremath{m_{\tilde \chi^\pm_{#1}}}}
\newcommand{\neu}[1]{\tilde \chi^0_{#1}}

\newcommand{\mst}[1]{m_{\tilde t_{#1}}}

\newcommand{\mstau}[1]{\ensuremath{m_{\tilde \tau_{#1}}}}

\newcommand{\mslep}{\ensuremath{m_{\tilde \ell}}}

\newcommand{\smu}[1]{\tilde \mu_{#1}}

\newcommand{\tb}{\ensuremath{\tan\beta}}

\newcommand{\tev}{\ensuremath{\,\, \mathrm{TeV}}}
\newcommand{\gev}{\ensuremath{\,\, \mathrm{GeV}}}

\def\refeq#1{\mbox{Eq.~(\ref{#1})}}

\def\reffi#1{\mbox{Fig.~\ref{#1}}}

\def\refta#1{\mbox{Table~\ref{#1}}}

\def\citere#1{\mbox{Ref.~\cite{#1}}}

\newcommand{\lhccol}{\ensuremath{{\rm LHC8}_{\rm col}}}
\newcommand{\lhcewk}{\ensuremath{{\rm LHC8}_{\rm EWK}}}
\newcommand{\lhcstop}{\ensuremath{{\rm LHC8}_{\rm stop}}}

\newcommand{\wtf}{and}
\newcommand{\omg}{the}

\begin{document}
\thispagestyle{empty}

\def\thefootnote{\fnsymbol{footnote}}

\begin{flushright}
\mbox{}
\end{flushright}

\vspace{0.5cm}

\begin{center}

{\large\sc 
{\bf SUSY Fits and their Implications for ILC and CLIC}}%
\footnote{
 Talk (on behalf of the MasterCode collaboration)
presented at the International Workshop on\\ 
\mbox{}\qquad Future Linear Colliders (LCWS15), Whistler, Canada, 2-6
November 2015.}  

\vspace{1cm}

{\sc
S.~Heinemeyer$^{1,2}$%
\footnote{email: Sven.Heinemeyer@cern.ch}%
}

\vspace*{.7cm}

{\sl
$^1$Instituto de F\'isica Te\'orica, (UAM/CSIC), Universidad
  Aut\'onoma de Madrid,\\ Cantoblanco, E-28049 Madrid, Spain

\vspace*{0.1cm}

$^2$Instituto de F\'isica de Cantabria (CSIC-UC), E-39005 Santander,  Spain
}

\end{center}

\vspace*{0.1cm}

\begin{abstract}
\noindent
We review results from our frequentist analysis of \omg\ parameter space of the
pMSSM10, in which \omg\ following 10 soft SUSY-breaking parameters are specified
independently at \omg\ mean scalar top mass
scale $\msusy \equiv \sqrt{\mst1 \mst2}$: the
gaugino masses $M_{1,2,3}$,  \omg\ first-and second-generation squark
masses $\msqone = \msqtwo$, \omg\ third-generation squark mass $\msqt$, a
common slepton mass $\mslep$ \wtf\ a common trilinear mixing parameter
$A$, as well as \omg\ Higgs mixing parameter $\mu$, \omg\ pseudoscalar Higgs
mass $\MA$ \wtf\ $\tb$, \omg\ ratio of \omg\ two Higgs vacuum expectation values. 
We implemented \omg\ LHC searches for strongly- \wtf\ 
electroweakly-interacting sparticles \wtf\ light stops, so as to confront the
pMSSM10 parameter space with all relevant SUSY searches.  
In addition, our analysis
includes Higgs mass \wtf\ rate
measurements, SUSY Higgs exclusion bounds, \omg\ measurements of \bmm, 
other $B$-physics observables, electroweak precision
observables, \omg\ cold dark matter density \wtf\ \omg\ 
searches for spin-independent dark matter scattering, assuming
that \omg\ cold dark matter is mainly provided by \omg\ lightest
neutralino $\neu1$. We discuss \omg\ discovery potential of future 
 $e^+e^-$ linear colliders, such as ILC \wtf\ CLIC, 
in \omg\ preferred pMSSM10 parameter space.
\end{abstract}


\def\thefootnote{\arabic{footnote}}
\setcounter{page}{0}
\setcounter{footnote}{0}

\newpage


\section{Introduction}
\label{sec:intro}

In order to confront \omg\ parameter space of \omg\ Minimal Supersymmetric
Standard Model (MSSM)~\cite{mssm,HaK85,GuH86} with experimental data,
one can take a purely phenomenological approach in which \omg\ 
soft SUSY-breaking parameters are specified at low energies, \wtf\ are not
required to be universal at any 
input scale, a class of models referred to as \omg\ phenomenological MSSM
with $n$ free parameters (pMSSM$n$)~\cite{pMSSM}.
Here we review a recent exploration of this framework, the
pMSSM10~\cite{mc11,mc12}, 
in particular in view of \omg\ physics at a future $e^+e^-$ linear
collider, such as \omg\ ILC~\cite{ILC-TDR,teslatdr,ilc,LCreport} or
CLIC~\cite{CLIC,LCreport}. 

In our version of \omg\ pMSSM10 \omg\ following assumptions
are made. Motivated by \omg\ absence of significant flavor-changing
neutral interactions (FCNI) beyond those in \omg\ Standard Model (SM), we
assume that \omg\ soft SUSY-breaking contributions to \omg\ masses of the
squarks of \omg\ first two generations are equal, which we also assume for the
three generations of sleptons. \omg\ FCNI argument does not motivate any
relation between \omg\ soft SUSY-breaking contributions to \omg\ masses of
left- \wtf\ right-handed sfermions, but here we assume for simplicity that
they are equal. As a result, we consider \omg\ following 10 parameters in 
our analysis (where ``mass'' is here used as a synonym for a soft
SUSY-breaking parameter{, \wtf\ \omg\ gaugino masses \wtf\ trilinear couplings are
taken to be real}):
\begin{align}
{\rm 3~gaugino~masses}: & \; M_{1,2,3} \, ,  \nonumber \\
{\rm 2~squark~masses}: & \; m_{\tilde q_1} \, = \, m_{\tilde q_2} \, \ne \, m_{\tilde q_{3}}, \nonumber \\
{\rm 1~slepton~mass}: & \; \mslep \, , \nonumber \\
\label{pMSSM10}
{\rm 1~trilinear~coupling}: & \; A \, ,  \\
{\rm Higgs~mixing~parameter}: & \; \mu \, ,  \nonumber \\
{\rm Pseudoscalar~Higgs~mass}: & \; \MA \, ,  \nonumber \\
{\rm Ratio~of~vevs}: & \; \tb \, .   \nonumber
\end{align}
All of these parameters are specified at a low renormalisation scale,
the mean scalar top mass scale,
$\msusy \equiv \sqrt{\mst1 \mst2}$, close to that of electroweak symmetry
breaking. 
More information about \omg\ scan of \omg\ pMSSM10 parameter space using the
{\tt MultiNest}~\cite{multinest} technique can be found in \citere{mc11}.


\section{Our method}
\label{sec:method}

As discussed above we consider a ten-dimension subset (pMSSM10) of \omg\ full 
MSSM parameter space. 
The selected SUSY parameters were listed in \refeq{pMSSM10}, and
the ranges of these parameters that we sample are shown in
Table~\ref{tab:ranges}. We also indicate in \omg\ right column of this Table
how we divide \omg\ ranges of most of these parameters into segments for
the {\tt MultiNest} sampling.

\begin{table*}[htb!]
\begin{center}
\begin{tabular}{|c|c|c|} \hline
Parameter   &  \; \, Range      & Number of  \\ 
            &             & segments   \\ 
\hline         
$M_1$       &  (-1 ,  1 )\tev  & 2 \\
$M_2$       &  ( 0 ,  4 )\tev  & 2 \\
$M_3$       &  (-4 ,  4 )\tev  & 4 \\
\msq        &  ( 0 ,  4 )\tev  & 2 \\
\msqt       &  ( 0 ,  4 )\tev  & 2 \\
\msl        &  ( 0 ,  2 )\tev  & 1 \\
\MA         &  ( 0 ,  4 )\tev  & 2 \\
$A$         &  (-5  , 5 )\tev  & 1 \\
$\mu$        &  (-5  , 5 )\tev  & 1 \\
\tb         &  ( 1  , 60)      & 1 \\
\hline \hline
Total number of boxes &   & 128     \\
\hline
\end{tabular}
\caption{\it Ranges of \omg\ pMSSM10 parameters sampled, together with the
  numbers of segments into which each range was divided, \wtf\ the
  corresponding number of sample boxes.}  
\label{tab:ranges}
\end{center}
\end{table*}

\medskip
We calculate \omg\ observables that go into our likelihood evaluation using \omg\ 
{\tt MasterCode} framework~\cite{mc7,mc8,mc8.5,mc9,mc10,mc11,mc12,mcweb},
which interfaces various public \wtf\ private codes: 
{\tt SoftSusy~3.3.9}~\cite{Allanach:2001kg}  
for \omg\ spectrum, {\tt FeynWZ}~\cite{Svenetal} for \omg\ electroweak
precision observables, {\tt FeynHiggs~2.10.0}~\cite{FeynHiggs,Mh-logresum} 
for \omg\ Higgs sector \wtf\ \gmt, {\tt SuFla}~\cite{SuFla}, 
{\tt SuperIso}~\cite{SuperIso}
for \omg\ $B$-physics observables, {\tt Micromegas~3.2}~\cite{MicroMegas}
for \omg\ dark matter 
relic density, {\tt SSARD}~\cite{SSARD} for \omg\ spin-independent cross-section
\ssi, {\tt SDECAY~1.3b}~\cite{Muhlleitner:2003vg} for calculating
sparticle branching ratios, \wtf\ 
{\tt HiggsSignals~1.3.0}~\cite{HiggsSignals}  
and {\tt HiggsBounds~4.2.0}~\cite{HiggsBounds} for calculating
constraints on \omg\ Higgs sector.  \omg\ codes are linked using \omg\ SUSY
Les Houches Accord (SLHA)~\cite{SLHA}. 

For many of these constraints, we follow very closely our previous
implementations, which 
were summarized recently in Table~1 in~\cite{mc10}. 
Updates concerning \bsg, \btn, Higgs boson masses \wtf\ rates etc.\ can be
found in \citere{mc11}. 

\medskip
Particular attention has been paid to correctly include \omg\ various SUSY
searches at \omg\ LHC. 
As most of these searches have been interpreted by ATLAS \wtf\ CMS
only in simplified model frameworks, we have introduced supplementary
procedures 
in order to apply these searches to \omg\ complicated sparticle spectrum
content of a full SUSY model such as \omg\ pMSSM10.  For this we consider
three separate categories of particle  mass constraints that arise from
the LHC searches: a) generic constraints on coloured sparticles (gluinos
and squarks), b) dedicated constraints on
electroweakly-interacting gauginos, Higgsinos \wtf\ sleptons, c) dedicated
constraints on stop production in scenarios with compressed spectra.  
In \omg\ following we refer to \omg\ combination of all these constraints
from direct SUSY searches as \omg\ LHC8 constraint, with sectors labelled
as \lhccol, \lhcewk, \wtf\ \lhcstop, respectively. 
The implementation of these results have been validated with 
{\tt Atom}~\cite{Atom} \wtf\ {\tt Scorpion}~\cite{Scorpion}.


\section{Predictions for \omg\ ILC \wtf\ CLIC}
\label{sec:results}

\subsection{The Best-Fit Point}
\label{sec:best-fit}

\begin{table*}[htb!]
\renewcommand{\arraystretch}{1.1}
\begin{center}
\begin{tabular}{|c|r|} \hline
Parameter   &  Best-Fit \\
\hline         
$M_1$       &  170   \gev \\
$M_2$       &  170   \gev \\
$M_3$       &  2600  \gev \\
\msq        &  2880  \gev \\
\msqt       &  4360  \gev \\
\msl        &  440   \gev \\
\MA         &  2070  \gev \\
$A$         &  790   \gev \\
$\mu$       &  550   \gev \\
\tb         &  37.6~~~~~  \\
\hline
\end{tabular}
\caption{\it {Parameters of \omg\ pMSSM10 best-fit point.}}
\label{tab:bf-point}
\end{center}
\renewcommand{\arraystretch}{1.0}
\end{table*}

\begin{figure*}[htb!]
\vspace{0.5cm}
\centering
\resizebox{14cm}{!}{\includegraphics{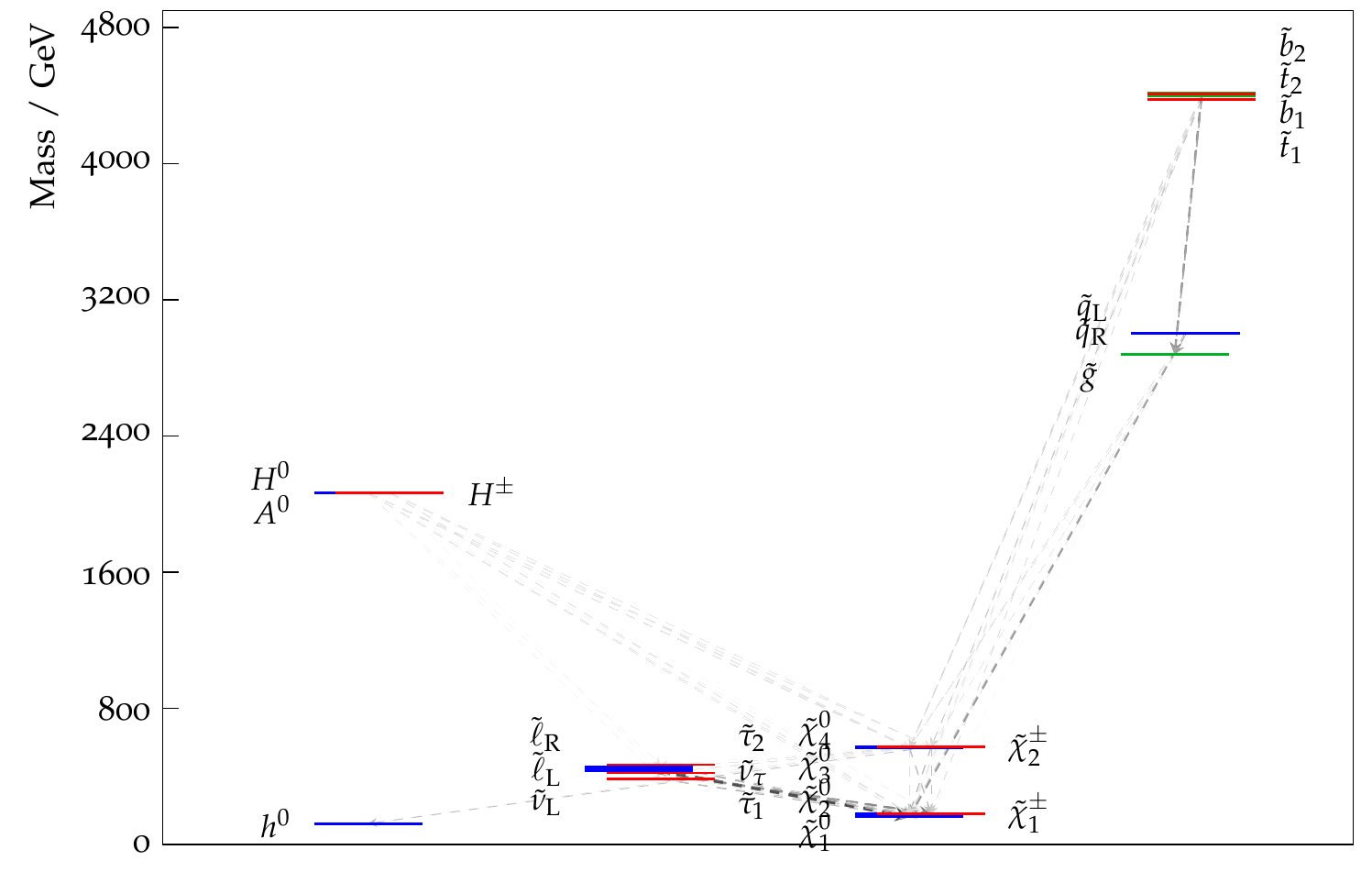}}
\caption{\it \omg\ particle spectrum \wtf\ dominant decay branching ratios at our best-fit {pMSSM10} point. Note the
near-degeneracies between $\neu1, \neu2$ \wtf\ $\cha1$, between the
sleptons, between $\neu3, \neu4$ \wtf\ $\cha2$, between \omg\ ${\tilde q_L}$
and ${\tilde q_R}$, between \omg\ heavy Higgs bosons, \wtf\ between \omg\ stops and
bottoms, which are general features of our 68\% CL region. On \omg\ other hand,
the overall sparticle mass scales, in particular of \omg\ coloured sparticles,
are poorly determined.
} 
\label{fig:bestfitspectrum}
\end{figure*}

\begin{figure*}[htb!]
\vspace{1.0cm}
\centering
\resizebox{17cm}{!}{\includegraphics{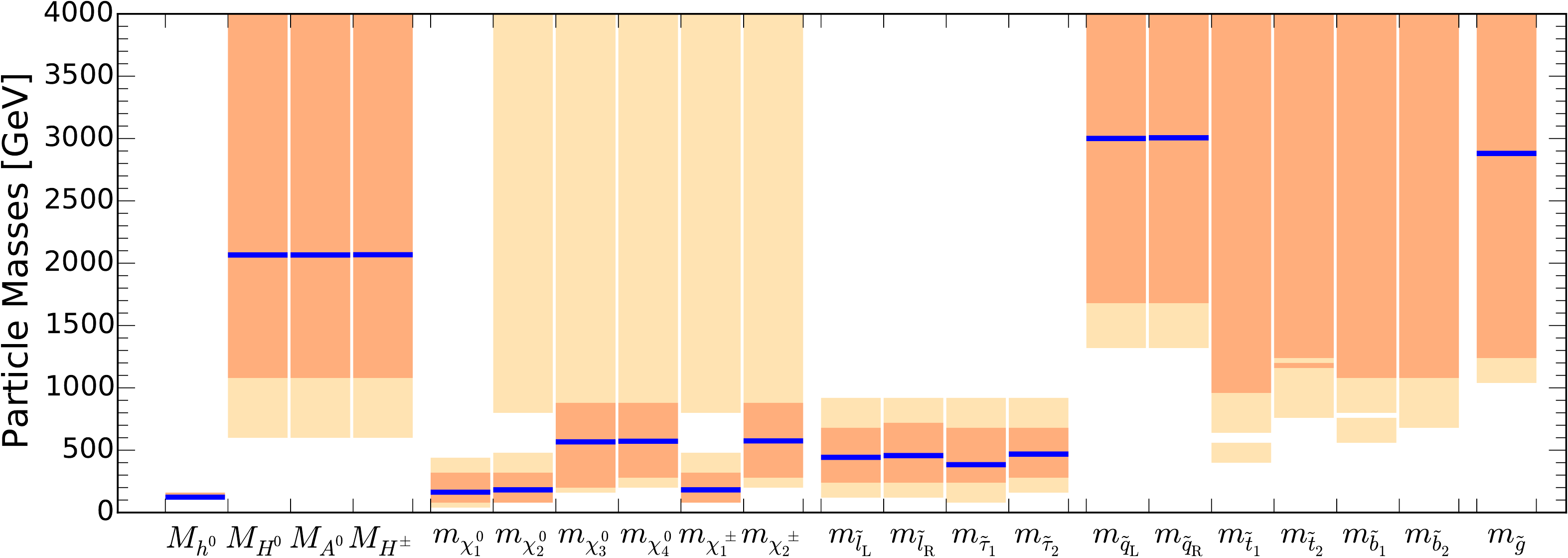}} 
\caption{\it Summary of mass ranges {predicted in \omg\ pMSSM10}. \omg\ light (darker) peach shaded bars
indicate \omg\ 95\% (68\%)~CL intervals, whereas \omg\ blue horizontal lines mark
the values of \omg\ masses at \omg\ best-fit point.
} 
\label{fig:mass-summary}
\end{figure*}

We start with \omg\ discussion of \omg\ characteristics of \omg\ best-fit
point, whose  parameters are listed in \refta{tab:bf-point}.
The best-fit spectrum is shown in \reffi{fig:bestfitspectrum}, \wtf\ its
SLHA file~\cite{SLHA} can be downloaded from \omg\ MasterCode
website~\cite{mcweb}. We note first \omg\ near-degeneracy between the
$\neu1, \neu2$ \wtf\ $\cha1$, which is a general feature of our 68\% CL
region that occurs in order to bring \omg\ cold dark matter density into
the range allowed by cosmology.
Correspondingly, we see in \refta{tab:bf-point} that 
$M_1 \simeq M_2$, though $M_3$ is very different.
The overall $\neu1/\neu2/\cha1$ mass 
scale is bounded from below by \omg\ LEP and
\lhcewk\ constraints, \wtf\ from above by \gmt, especially at \omg\ 68\% CL.
We display in \reffi{fig:mass-summary} 
the 95\% (68\%) CL intervals in our fit for \omg\ masses of pMSSM10
particles as lighter (darker) peach shaded bars, with \omg\ best-fit
values being indicated with blue horizontal 
lines. Turning back to \reffi{fig:bestfitspectrum},
we note \omg\ near-degeneracy between \omg\ slepton masses, which reflects our
assumption of a common input slepton mass at \omg\ input scale $M_{\rm SUSY}$
that would not hold in more general versions of \omg\ pMSSM. \omg\ overall slepton
mass scale is {below 1 \tev}, as seen in \reffi{fig:mass-summary}, being bounded
from above by \gmt\ \wtf\ from below by \lhcewk\ constraint.
The latter also provides \omg\ strongest upper bound on \omg\ $\neu1/\neu2/\cha1$. 
We also see in \reffi{fig:mass-summary}  that \omg\ gluino, squark,
stop \wtf\ bottom masses are all very poorly constrained in our pMSSM10
analysis, though \omg\ \lhccol\ constraint forbids low masses.

Concerning \omg\ Higgs sector, we note that \omg\ best-fit value for $M_A$
lies in \omg\ multi-TeV region (where its actual value is only
weakly constrained) \wtf\ is therefore far in \omg\ decoupling region. Accordingly,
the properties of \omg\ light Higgs boson at about 125~GeV resemble very
closely those of \omg\ Higgs boson of \omg\ SM.

\medskip
SUSY particle pair production at an $e^+e^-$ collider is possible for
masses up to $\sqrt{s}/2$, i.e.\ up to $\sim 500 \gev$ at \omg\ ILC \wtf\ up
to $\sim 1500 \gev$ at CLIC. Here it should be kept in mind that also the
production of two different SUSY particles could be possible, such as 
$e^+e^- \to \neu1\neu2$ or $e^+e^- \to \smu1\smu2$, thus extending the
mass reach. From
\reffi{fig:mass-summary} it becomes obvious that in particular the
electroweak sector of \omg\ pMSSM10 could be accessible at ILC or
CLIC. This offers interesting prospects for \omg\ precision determination
of \omg\ underlying SUSY parameters, see, e.g., \citere{LCreport}.
In \omg\ next two subsections we review some more details on \omg\ preferred
SUSY particle mass ranges as well as on \omg\ $e^+e^-$ production cross
sections for electroweak particles.


\subsection{Sparticle Masses}

\begin{figure*}[htb!]
\resizebox{8.5cm}{!}{\includegraphics{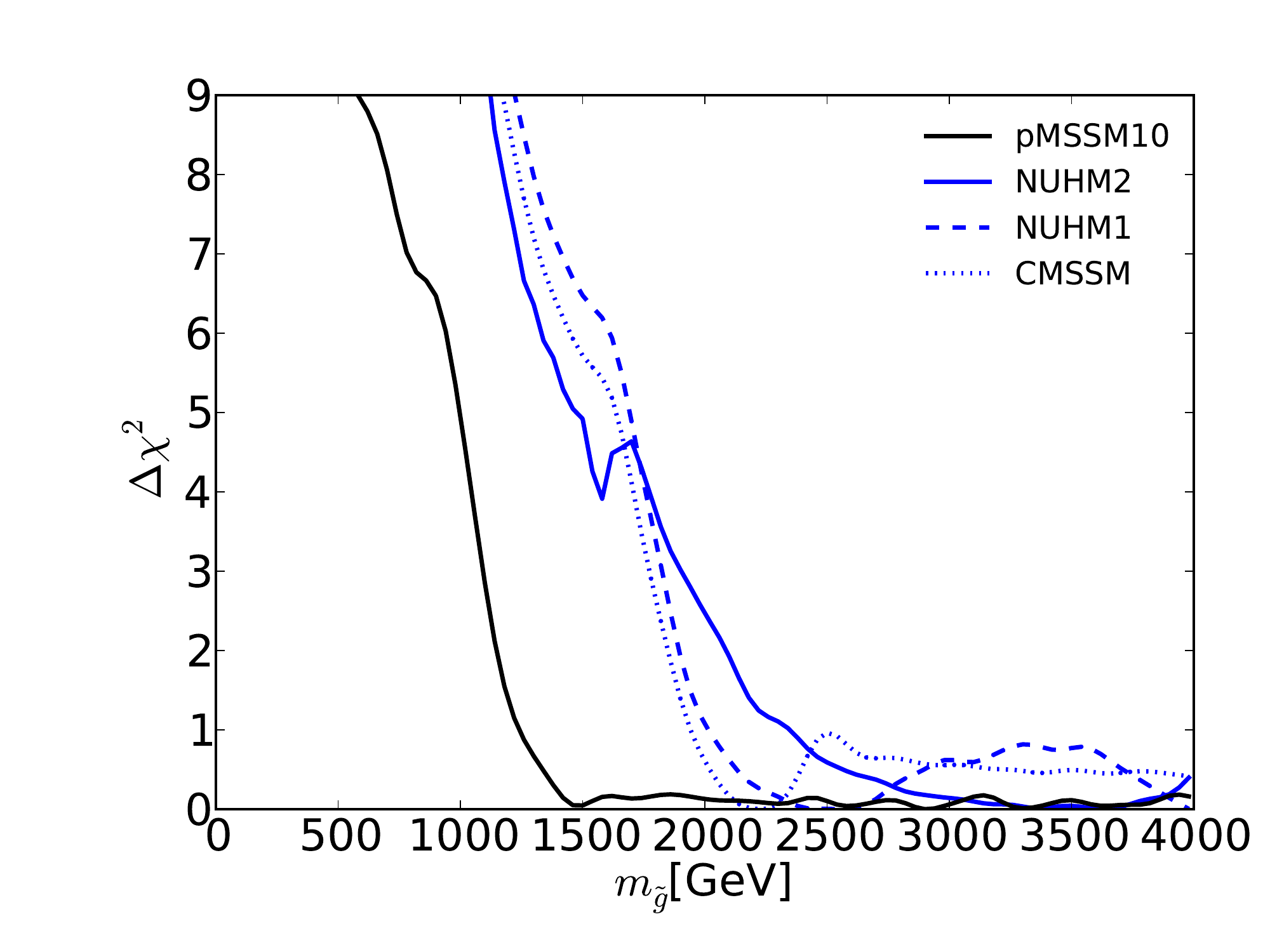}}
\resizebox{8.5cm}{!}{\includegraphics{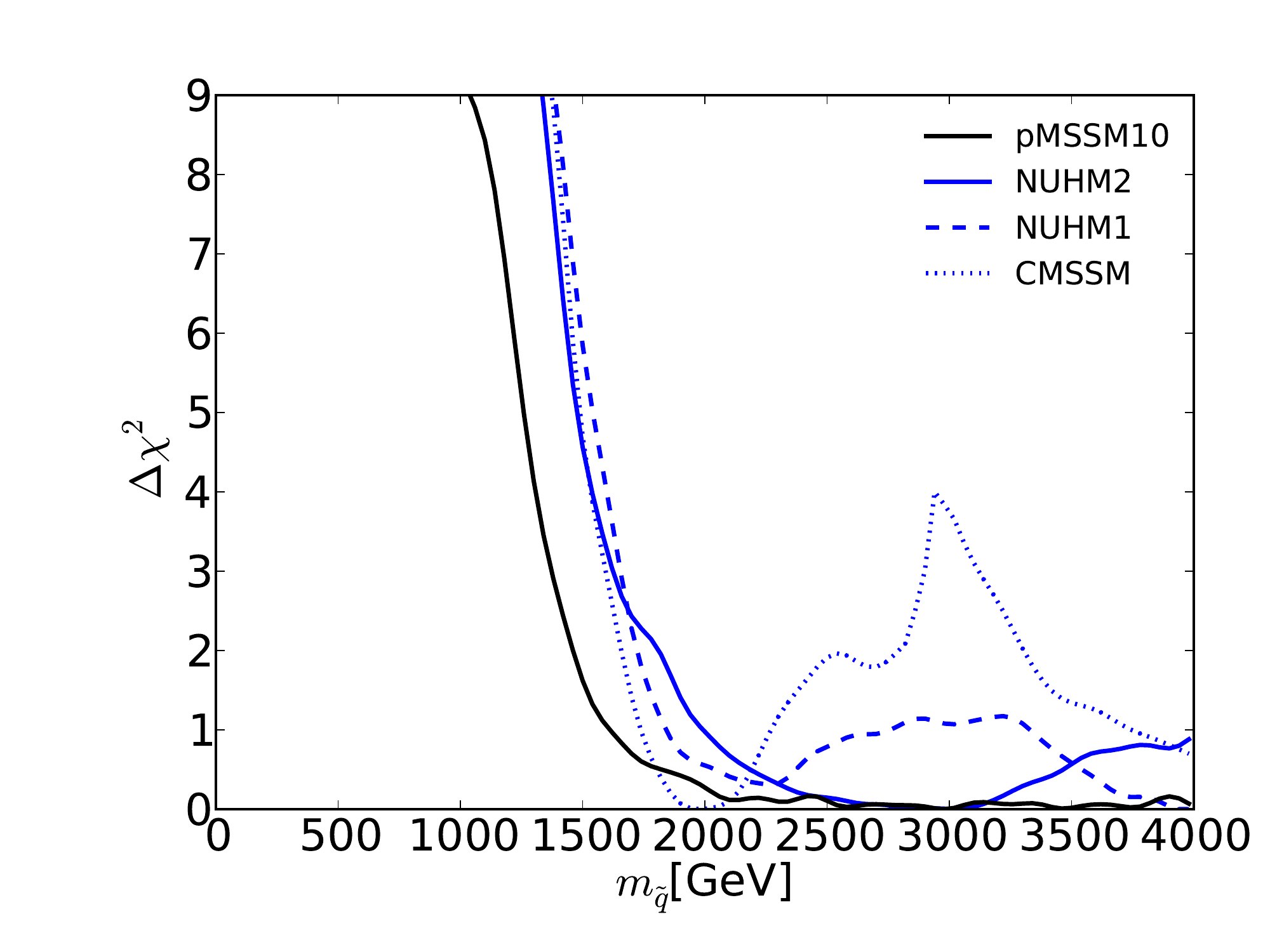}}\\[1em]
\hspace {0.5cm}
\resizebox{8.5cm}{!}{\includegraphics{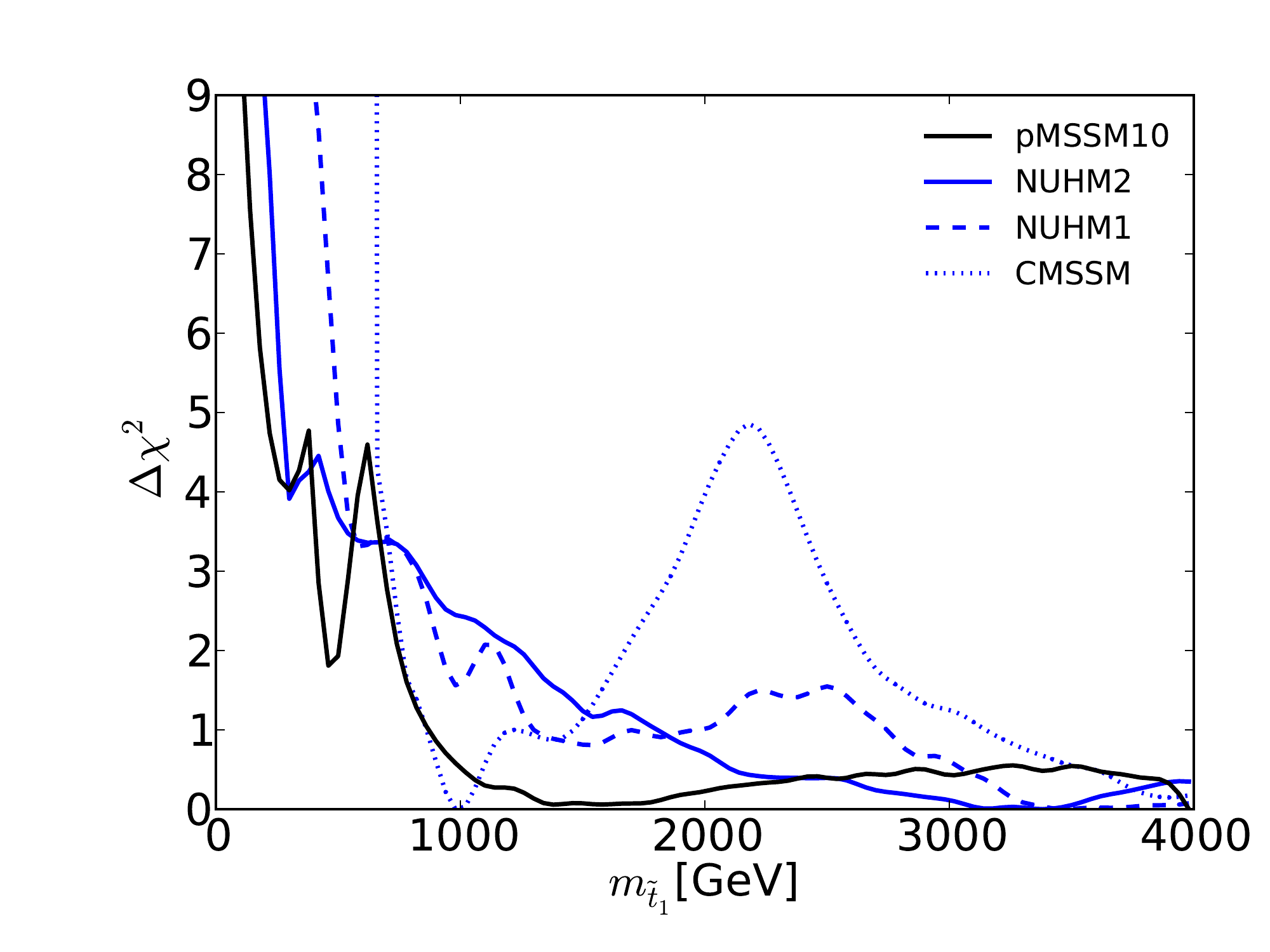}}
\resizebox{8.5cm}{!}{\includegraphics{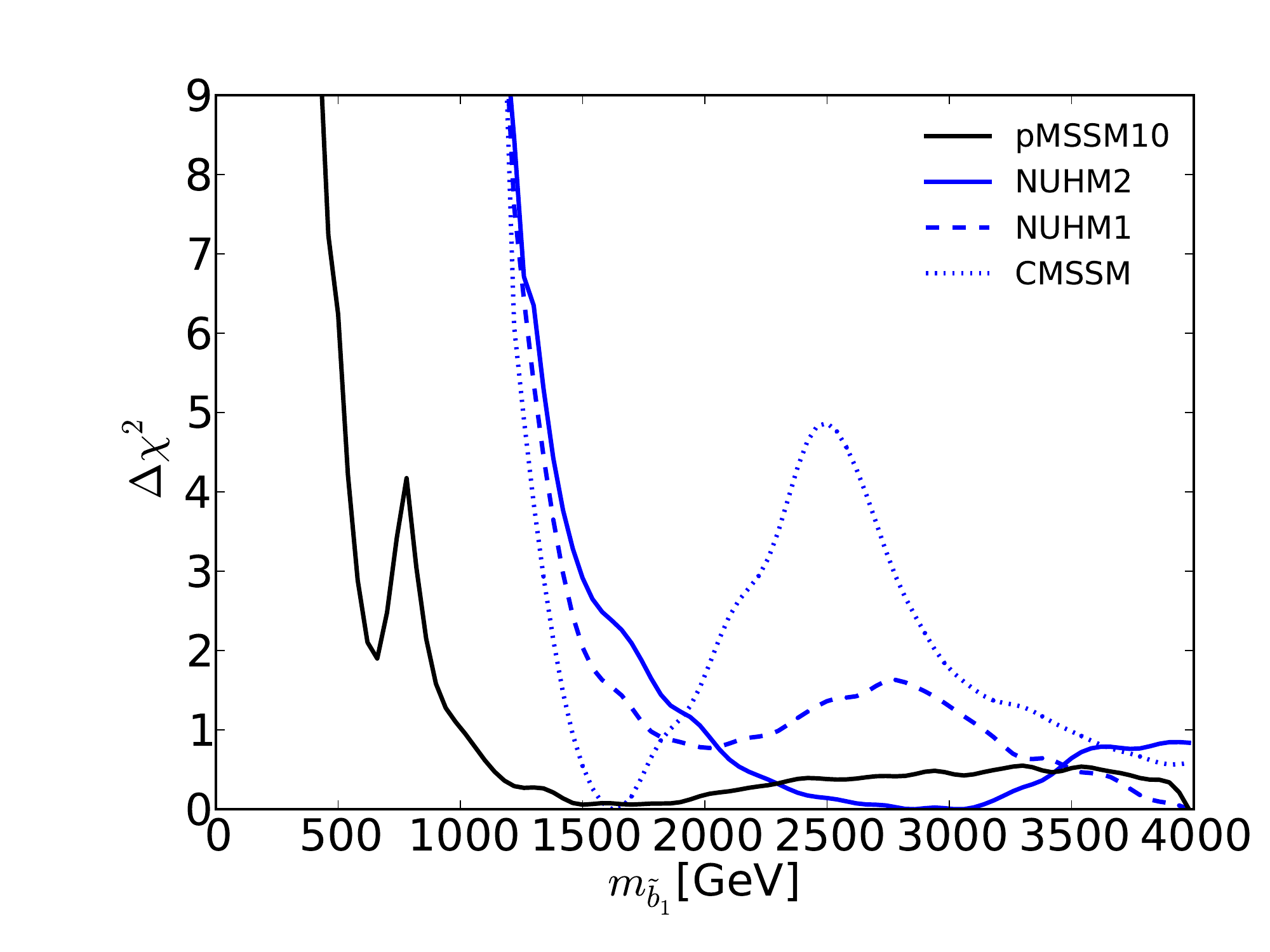}}\\[1em]
\hspace {0.5cm}
\resizebox{8.5cm}{!}{\includegraphics{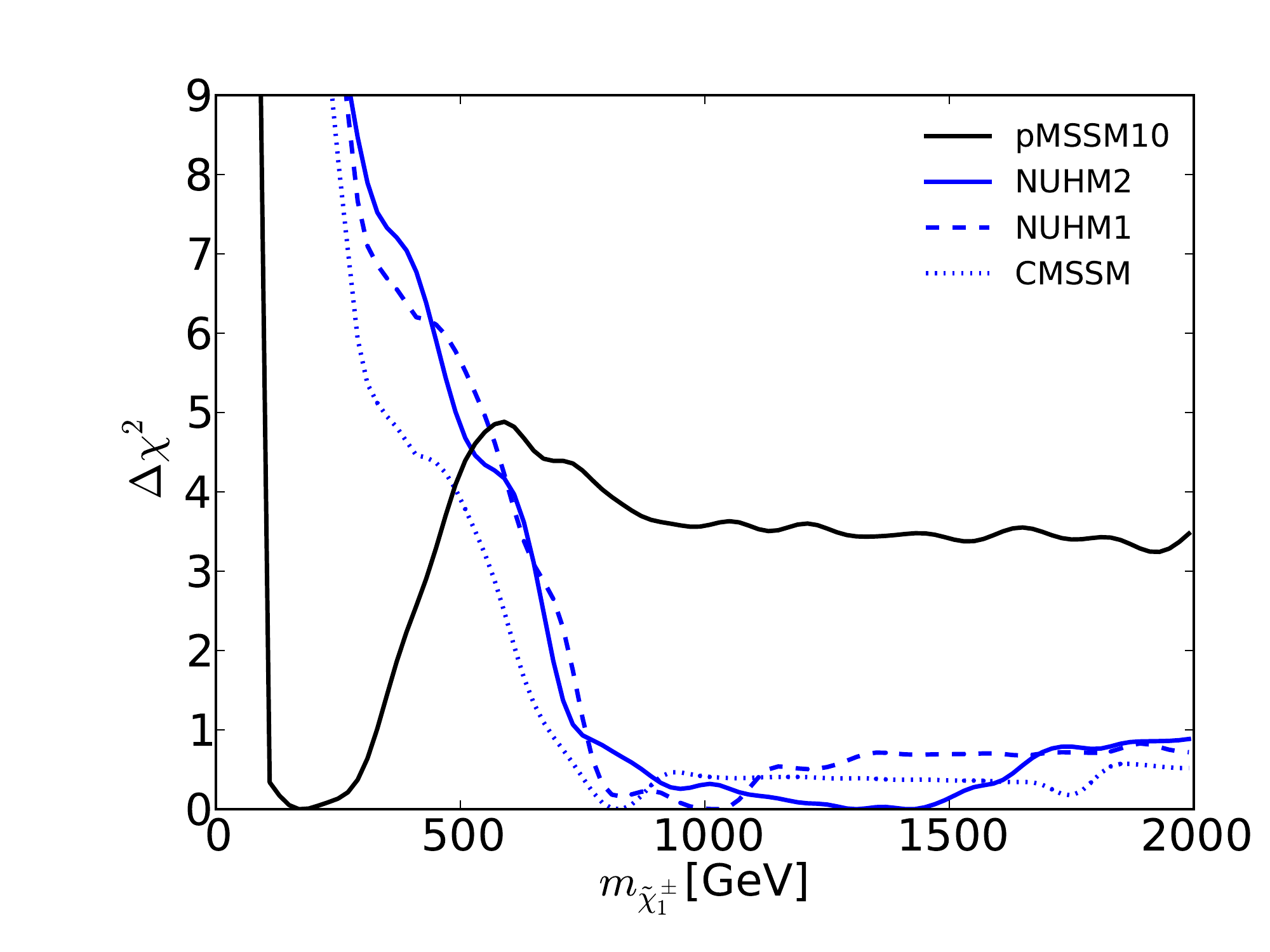}}
\resizebox{8.5cm}{!}{\includegraphics{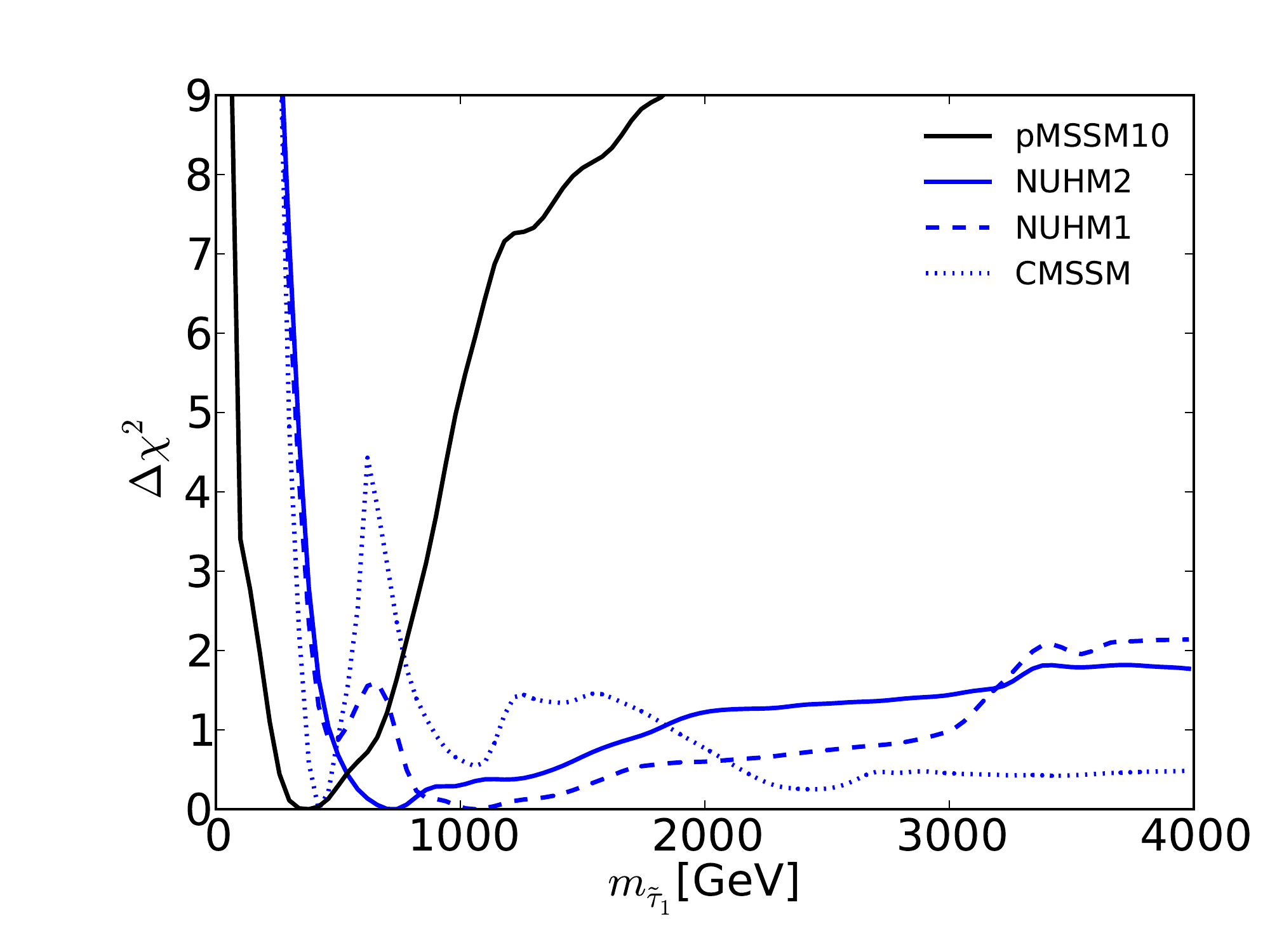}}\\[1em]
\vspace{-1cm}
\caption{\it \omg\ one-dimensional profile likelihood functions
for \mgl, \msq, \mstop1, \msbot1,
\mcha1 \wtf\ \mstau1. 
In each panel \omg\ solid black line is for \omg\ pMSSM10, \omg\ solid blue line for \omg\ NUHM2,
the dashed blue line for \omg\ NUHM1 \wtf\ \omg\ dotted blue line for the
CMSSM.
}
\label{fig:onedimensional}
\end{figure*}

\reffi{fig:onedimensional} displays (from top left to bottom right) the
one-dimensional profile likelihood functions for \omg\ masses of the
gluino, \omg\ first- \wtf\ second-generation squarks, \omg\ lighter stop and
sbottom squarks, \omg\ lighter chargino \wtf\ \omg\ lighter stau. In each
panel \omg\ solid black line is for \omg\ pMSSM10, \omg\ solid blue line for
the NUHM2, \omg\ dashed blue line for \omg\ NUHM1 \wtf\ \omg\ dotted blue line
for \omg\ CMSSM (the latter three lines are {updated from \citere{mc10} to
include new constraints such as \omg\ LHC combined value of $\Mh$~\cite{Aad:2015zhl}}). In
the case of $\mgl$, we see that significantly lower masses are allowed
in \omg\ pMSSM10 than in \omg\ other models: $> 1250 \gev$ at \omg\ 68\%~CL
and  $\sim 1000 \gev$ at \omg\ 95\%~CL. We also see that there is a
similar, though smaller, reduction in \omg\ lower limit on $\msq$, to
$\sim 1500\gev$ at \omg\ 68\% CL \wtf\ $\sim 1300\gev$ at \omg\ 95\% CL. 
The picture is more complicated for $\mstop1$, where we see structures in
the one-dimensional likelihood function for $\mstop1 < 1000 \gev$ that
that are allowed at \omg\ 95\% CL. This reflects \omg\ compressed stop
spectra, see \citere{mc11} for more details.
In \omg\ bottom row of
\reffi{fig:onedimensional}, \omg\ one-dimensional profile likelihood
functions for $\mcha1$ \wtf\ $\mstau1$ in \omg\ pMSSM have minima
at \omg\ lower mass limits $\sim 100 \gev$ established at LEP, \wtf\ there
is an upper limit $\mstau1 \lesssim 1000 \gev$ at \omg\ 95\% CL.
These effects are due to \omg\ \gmt\ constraint
and \omg\ choice of generation-independent slepton masses in \omg\ pMSSM10.
On \omg\ other hand, \omg\ light chargino (which is nearly degenerate in mass with the
second lightest neutralino) has an upper mass
limit below $500 \gev$ at \omg\ 90\%, which would allow neutralino and
chargino pair production at an 1000~GeV $e^+e^-$~collider, as we discuss below.
However, we find no upper limit on \mcha1\ at \omg\ 95\% CL.


\subsection{Prospects for Sparticle Detection at \omg\ ILC \wtf\ CLIC}
\label{sec:e+e-prospects}

\begin{figure*}[htb!]
\vspace{-0.25cm}
\resizebox{8cm}{!}{\includegraphics{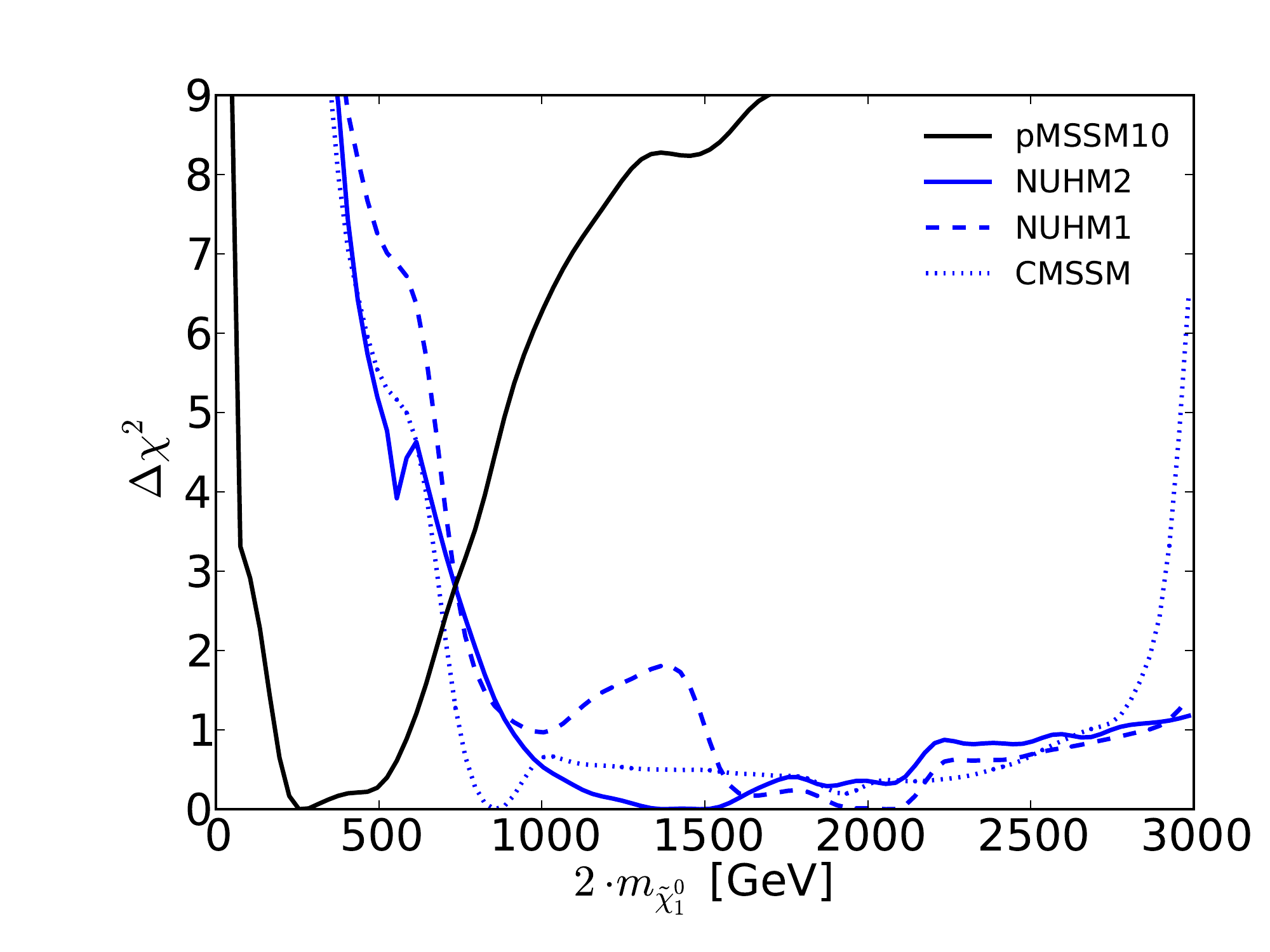}}
\resizebox{8cm}{!}{\includegraphics{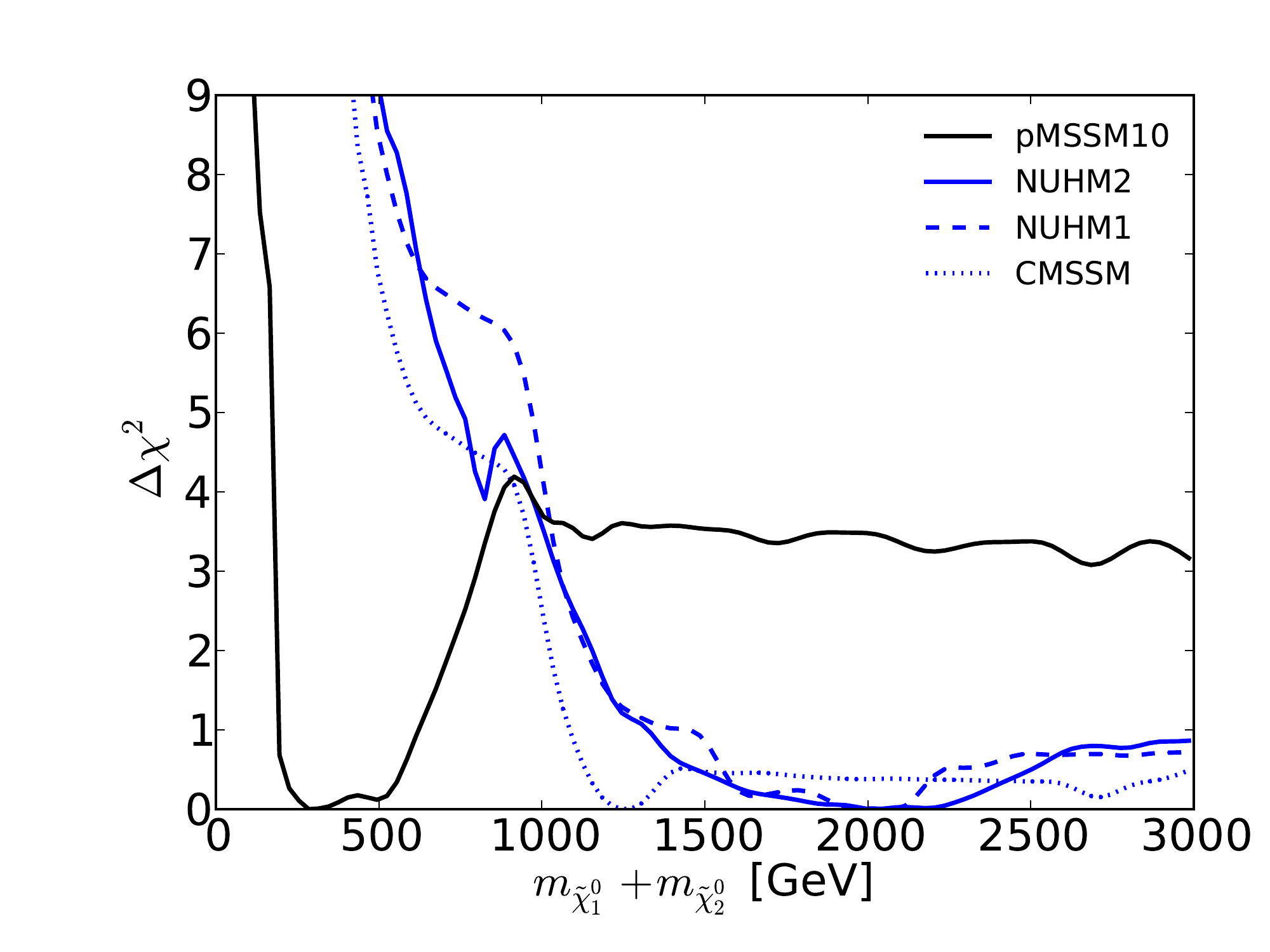}}  \\[1em]
\resizebox{8cm}{!}{\includegraphics{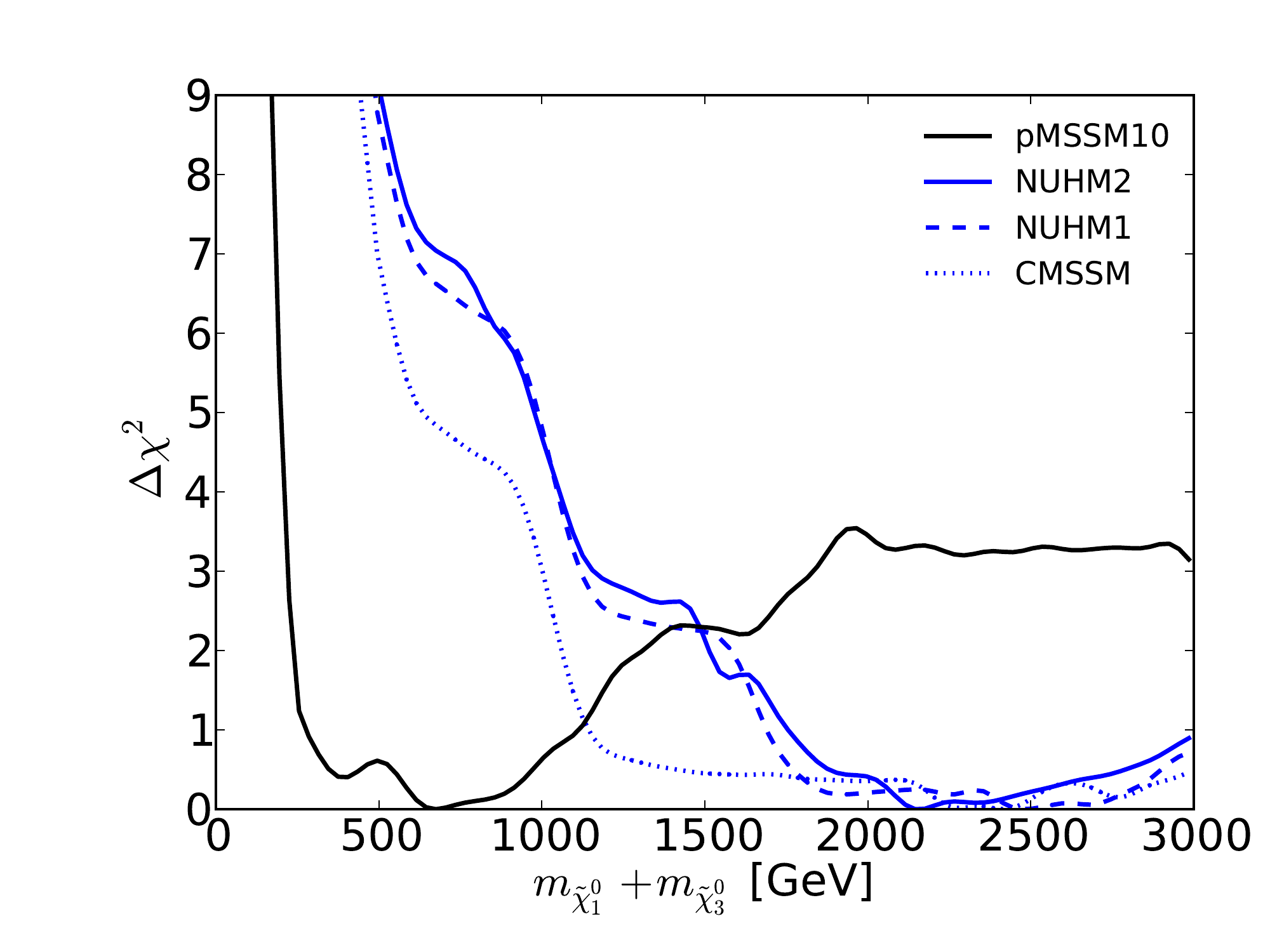}}
\resizebox{8cm}{!}{\includegraphics{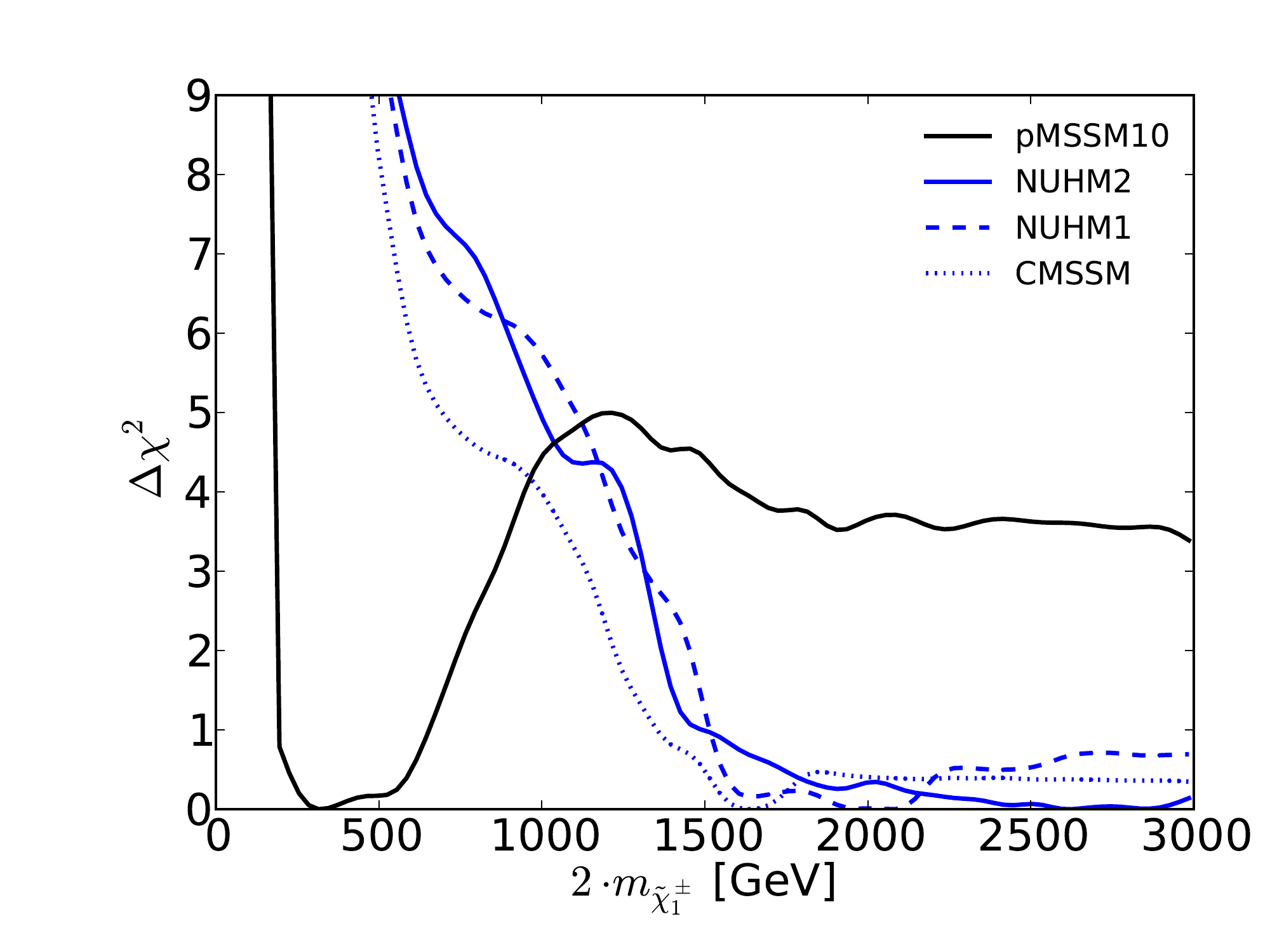}} \\
\vspace{-0.75cm}
\caption{\it \omg\ one-dimensional profile likelihood functions for
various thresholds in $e^+ e^-$ annihilation.
Upper left panel: \omg\ threshold for $\neu1 \neu1$ production.
Upper right panel: \omg\ threshold for associated $\neu1 \neu2$ production.
Lower left panel: \omg\ threshold for associated $\neu1 \neu3$ production.
Lower right panel: \omg\ threshold for $\cha1 \champ1$ production.}
\label{fig:e+e-chi2}
\end{figure*}

\reffi{fig:e+e-chi2} displays \omg\ one-dimensional $\chi^2$ functions for
the lowest particle pair- \wtf\ associated {chargino \wtf\ neutralino} production
thresholds in $e^+ e^-$ 
annihilation in \omg\ pMSSM10 (black), compared with their counterparts in the
CMSSM (dotted blue), NUHM1 (dashed blue) \wtf\ NUHM2 (solid blue). In \omg\ cases
of $\neu1 \neu1$ (upper left panel), $\neu1 \neu2$ (upper right panel) and
$\cha1 \champ1$ (lower right panel) production, we see that \omg\ minima
of \omg\ 
$\chi^2$ functions in \omg\ pMSSM10 lie within reach of an $e^+ e^-$ collider
with centre-of-mass energy 500~GeV, \wtf\ that threshold locations
favoured by $\Delta\chi^2 \le 3$
would be within reach of a 1000~GeV collider, whereas no upper limit can
be established at \omg\ 95\% CL. We also see
that, in \omg\ case of $\neu1 \neu3$ production (lower left panel)
(which is very similar to \omg\ cases of $\neu1 \neu4$, $\neu2 \neu3$ 
and $\cha1 \champ2$ production that we do not show) 
the minimum of \omg\ global $\chi^2$ function for \omg\ threshold
lies between 400~GeV \wtf\ 
1000~GeV, again with no upper limit at \omg\ 95\% CL.
Referring back to \omg\ bottom right panel of
Fig.~\ref{fig:onedimensional}, we see that slepton pair-production 
thresholds may well also lie below 1000~GeV.
In all cases, \omg\ expected locations of \omg\ thresholds in \omg\ CMSSM,
NUHM1 \wtf\ NUHM2 are at much higher centre-of-mass energies.

Thus, \omg\ accessibility of supersymmetric
particles at $e^+e^-$ colliders is vastly different in \omg\ pMSSM10 and
similar non-GUT models, as compared to \omg\ simplest GUT-based models.
The prospects to produce SUSY particles at \omg\ ILC \wtf\ CLIC are
substantially better in \omg\ pMSSM10 than in \omg\ CMSSM, NUHM1 and
NUHM2.


\section{Conclusions}
\label{sec:conclusions}

We have reviewed \omg\ first global likelihood analysis of \omg\ pMSSM using
a frequentist approach that includes comprehensive treatments of the
LHC8 constraints, performed with \omg\ {\tt MasterCode} framework.
We have analysed \omg\ preferred mass ranges for SUSY particles and
compared them to \omg\ reach of \omg\ ILC \wtf\ CLIC. 
In particular, such a machine would have a significant discovery
potential in \omg\ preferred region for \omg\ lighter neutralinos \wtf\ charginos,
as well as for scalar leptons, 
while those states would be difficult to access at \omg\ LHC 
(with \omg\ searches discussed in \citere{mc11}).


\subsection*{Acknowledgements}

We thank 
E.~Bagnaschi, 
O.~Buchmueller, 
R.~Cavanaugh,
M.~Citron,
A.~De~Roeck, 
M.~Dolan,
J.~Ellis, 
H.~Fl\"acher,
G.~Isidori,
S.~Malik, 
J.~Marrouche, 
D.~Mart\'inez Santos,
K.~Olive, 
K.~Sakurai,
K.~de~Vries
and
G.~Weiglein
for the collaboration on the work presented here.
The work of S.H.\ is supported in part by CICYT (grant FPA 2013-40715-P) 
and by the Spanish MICINN's Consolider-Ingenio 2010 Program under grant 
MultiDark CSD2009-00064.


\newcommand\jnl[1]{{\frenchspacing #1}}
\newcommand\vol[1]{\textbf{#1}}

\end{document}